\documentclass[a4paper]{jpconf}
\usepackage{graphicx}
\usepackage{latexsym}
\usepackage{dcolumn}
\usepackage{bm}
\usepackage{amsfonts, amsmath, amsthm, amssymb}
\usepackage{cite}
\usepackage{hyperref}

\newcommand{\be}{\begin{equation}}
\newcommand{\ee}{\end{equation}}
\newcommand{\bea}{\begin{eqnarray}}
\newcommand{\eea}{\end{eqnarray}}
\newcommand{\bse}{\begin{subequations}}
\newcommand{\ese}{\end{subequations}}

\begin{document}

\title{Dynamics of fluctuations in the Gaussian model with dissipative Langevin Dynamics}

\author{F Corberi$^{1,2}$, O Mazzarisi$^{1}$, A Gambassi$^{3,4}$}

\address{$^{1}$ Dipartimento di Fisica ``E.~R. Caianiello'', Universit\`a  di Salerno, via Giovanni Paolo II 132, 84084 Fisciano (SA), Italy.}
\address{$^{2}$ INFN, Gruppo Collegato di Salerno, 
and CNISM, Unit\`a di Salerno, Universit\`a  di Salerno, via Giovanni Paolo II 132, 84084 Fisciano (SA), Italy.}
\address{$^{3}$ SISSA - International School for Advanced Studies, via Bonomea 265, 34136 Trieste, Italy.}
\address{$^{4}$ INFN, Sezione di Trieste, via Bonomea 265, 34136, Trieste, Italy.}

\ead{onofrio.mazzarisi@gmail.com}

\begin{abstract}

We study the dynamics of the fluctuations of the variance $s$ of the order parameter of the Gaussian model, 
following a temperature quench of the thermal bath. 
At each time $t$, there is a critical value $s_c(t)$ of $s$
such that fluctuations with $s>s_c(t)$ are realized by condensed configurations of the systems, i.e.,
a single degree of freedom contributes macroscopically to $s$. This phenomenon, which is closely related
to the usual condensation occurring on average quantities, is usually referred to as {\it condensation of fluctuations}.
We show that the probability of fluctuations with $s<\inf_t [s_c(t)]$, associated
to configurations that never condense, after the quench converges
rapidly and in an adiabatic way towards the new equilibrium value.  
The probability of fluctuations with $s>\inf _t [s_c(t)]$, instead, displays a slow and more complex behavior, 
because the macroscopic population of the condensing degree of freedom is involved.

\end{abstract}

\section{Introduction} \label{intro}

Rare events in statistical systems play a relevant role in several fields of science~\cite{Hinrichsen00,Langer92,Touchette2009}.
Understanding these phenomena is therefore of paramount theoretical and practical importance.
The law of large numbers, which concerns only the average values of the observables of interest,
or the knowledge of the small fluctuations about the mean, which are regulated by the central limit theorem under 
certain assumptions, do not provide access to the interesting features emerging when large deviations set in. 
These rare fluctuations can play a significant role, for example, when they may drive a system
into an absorbing state~\cite{Hinrichsen00}.  

In order to investigate large deviations, consider a system which is described by
a set of (micro)variables \(\varphi = \{\varphi_i\}\),
where \(i\) is a generic label for the degrees of freedom,
with probability distribution \(P([\varphi],\bf t)\) depending on a set of control parameters \(\bf t\).
For Example, if the model system is a gas, $i$ labels the particles, 
$\varphi _i$ are the position and momenta of a molecule, and $\bf t$ represents temperature, volume etc.
The probability $P(S,\bf t)$ to observe a value $S$ of a collective variable \(\cal S [\varphi]\) (e.g., the energy in the
example) is
\begin{equation}
\label{eq:initialpdf}
P(S,\textbf {t})=\int  \mathcal D\varphi \ P([\varphi],{\bf t})\,\delta (S-{\cal S}[\varphi]).
\end{equation}
When one is able to access the full probability distribution \(P(S,\bf t)\),
all the information about the fluctuations are of course available. However, this is hardly the case, and a way
to learn something about fluctuations of \(\mathcal S[\varphi]\) beyond the central limit theorem
is to focus on large deviations in which the intensive variable
\(\mathcal S[\varphi]/V\) takes a certain value \(s\), where \(V\) is the system size~\cite{Touchette2009}.
Typically, the probability of the latter is suppressed as \(V\) increases, realizing the so called
large deviation principle (LDP).
A way to rationalize the LDP is the following:
consider Eq.~(\ref{eq:initialpdf}), using the Laplace representation of the delta function \(\delta (y)=(1/2\pi i)\int _{a-i\infty}^{a+i\infty} dz\, e^{-zy}\),
where $a$ is a real number such that the argument of the integral is analytic for ${\rm Re\,}z > a$, one can write
\begin{equation}
\label{eq:laplacerepresentation}
P(S,{\bf t})=\frac{1}{2\pi i}\int _{a-i\infty}^{a+i\infty} dz \ e^{-V\left [zs+\lambda (z,{\bf t})\right ]},
\end{equation}
where \(s=S/V\) is the intensive variable connected to \(S\)
and we introduced the scaled cumulant generating function 
\begin{equation}
\lambda(z,{\bf t} )=
-\frac{1}{V}\ln \int \mathcal D\varphi \ P([\varphi],{\bf t})e^{z{\cal S}[\varphi]}.
\end{equation}
For large \(V\), the integral in Eq.~(\ref{eq:laplacerepresentation}) can be evaluated via a saddle-point approximation
and thus
\begin{equation}
P(S,{\bf t})\sim e^{-VI(s,{\bf t})},
\label{largeDevPr}
\end{equation}
where we introduced the {\it rate function} \(I(s,{\bf t})\), which encodes all the information about the fluctuations
of the observable, defined by
\begin{equation}
I(s,{\bf t})\equiv z^*(s,{\bf t})s+\lambda(z^*(s,{\bf t}),{\bf t}),
\label{RateFunct}
\end{equation}
being $z^*(s,t)$ determined by the saddle-point condition \(\partial \lambda (z,\textbf t)/\partial z +s=0\).
Equation~(\ref{largeDevPr}) embodies the LDP.
The rate function is in general a non-concave function of \(s\), which has a minimum at where \(s\) equals its average value.
The LDP allows one to study exponentially (in $V$) rare fluctuations of $S$ of order $V$ which can display various 
interesting phenomena such as, e.g., singularities~\cite{Corberi19,Baek_2015,Filiasi_2014,Harris_2009,Gradenigo_2013,Gambassi2012,2019arXiv190406259P,Goold2018,Touchette2007,Touchette_2009,Bouchet_2012,Harris_2005,Szavits2014,Chleboun2010,Janas2016,Sasorov_2017,Majumdar_2014}. 
Calling $s_c$ one of these singular points, configurations of the system which realize fluctuations with $s<s_c$ or $s>s_c$ are qualitatively different, and this can be interpreted as a phase transition at the level of fluctuations~\cite{Corberi19}.

Large deviation theory has been successfully employed for studying stationary properties of both equilibrium and non-equilibrium stochastic processes~\cite{Touchette2009}. However, a clear understanding of the kinetics of large fluctuations is 
still lacking.
In this work we consider the fluctuations dynamics in the Gaussian model, a standard statistical mechanical model 
which may be regarded as the simplest Ginzburg-Landau theory for the description of the disordered phase of Ising-like 
systems~\cite{langer2000theory}. 
In this approach, 
the probability distribution of the order parameter variance $\mathcal S$, the observable we focus on, displays a singular 
point in $S=S_c$ both in and out of
equilibrium~\cite{Zannetti14,CORBERI2015,Corberi17,Zannetti14,Corberi19,Cagnetta17,Corberi_2015,Zannetti_2014,Corberi_2013,Corberi_2012,2019arXiv190508536C}. 
This fact is associated to the so-called {\it condensation of fluctuations}~\cite{Zannetti14},
a phenomenon whereby certain fluctuations are realized by a condensation mechanism in which a single
degree of freedom becomes macroscopically populated, similarly to what happens in usual condensation
phenomena, e.g., in the Bose gas. The difference is that in usual condensation the phenomenon is associated
to the typical behavior, whereas it occurs here only at the level of certain rare spontaneous fluctuations. 
Extending the work done in Ref.~\cite{2019arXiv190508536C}, we show that the dynamical properties of large deviations are non trivial in the presence of such a condensation. 

\section{The model} \label{themodel}

The Gaussian model~\cite{Goldenfeld92,chaikin_lubensky_1995} describes a real scalar field
$\varphi (\vec x)$ with Hamiltonian
\be
   {\cal H}[\varphi]=\frac{1}{2}\int _V d\vec x \left [(\nabla \varphi(\vec{x}) )^2
     +r\varphi ^2(\vec x)\right ],
   \label{ham}
\ee
where the parameter $r\ge 0$ is related to the equilibrium correlation length $\xi = r^{-1/2}$~\cite{Goldenfeld92}
and  \(V\) is the volume of the system in \(d\) spatial dimensions.
Here we focus on
a non-conserved order parameter (NCOP) dynamics, the so-called Model A~\cite{Hohenberg77}.
The time evolution of the field is thus given by the Langevin equation~\cite{Goldenfeld92}
\begin{equation}
\label{eq:Langevin}
\frac{\partial}{\partial t}\varphi(\vec x,t)=[\nabla^2-r]\varphi(\vec x,t)+\eta(\vec x,t),
\end{equation}
where $\eta (\vec x,t)$ is Gaussian white noise, due to a thermal bath in equilibrium at inverse temperature \(\beta\),
with zero average and correlations
\begin{equation}
\langle \eta (\vec x,t)\eta(\vec x',t')\rangle=2\beta^{-1}
\delta(\vec x -\vec x')\delta(t-t').
\end{equation}

The protocol we consider amounts at a temperature quench of the bath temperature, 
from $\beta_i =(k_B T_i)^{-1}$ to $\beta _f>\beta _i$, where \(k_B\) is
the Boltzmann constant, occurring at the time $t=0$.
The problem is linear and in Fourier space the solution reads
\begin{equation}
\varphi_{\vec{k}}(t)=\varphi_{\vec{k}}(0)e^{-\omega_{k}t}+\int_0^tdt'\ e^{-\omega_k(t-t')}\zeta_{\vec{k}}(t'),
\label{eq:solution_k-modes}
\end{equation}
where \(\varphi_{\vec k}(t)=\int_{V}d\vec x\varphi(\vec x,t)e^{i\vec k\cdot \vec x}\) are the Fourier components of 
the order parameter field (and similarly $\zeta _{\vec k}(t)$ for the noise $\eta (\vec x,t)$), and \(\omega_k:=k^2+r\).
The field correlator reads~\cite{Zannetti14}
\begin{equation}
\langle \varphi_{\vec{k}}(t)\varphi_{-\vec{k}}(t) \rangle =\langle\varphi_{\vec{k}}(0)\varphi_{-\vec{k}}(0)\rangle_0 e^{-2\omega_k t}+\frac{\beta_f^{-1} V}
        {\omega_k}(1-e^{-2\omega_k t}),
\end{equation}
with \(\langle...\rangle_0\) denoting average over the initial conditions.

In the scheme outlined in the previous section, the configuration \([\varphi]\) of the field at time \(t\), i.e., the collection of the 
\(k\)-modes, describes the microstate of the system. Due to the linear nature of the problem, the probability distribution 
\(P([\varphi],t)\) is Gaussian at all times, with
\begin{equation}
\label{eq:pdfmicrostates}
P([\varphi],t)=\prod_{\vec k}\mathcal Z^{-1}_{\vec k}(t)\exp\Bigl\{-\frac{\varphi_{\vec k}\varphi_{-\vec k}}{2\langle \varphi_{\vec{k}}(t)\varphi_{-\vec{k}}(t) \rangle}\Bigl\},
\end{equation}
where \(\mathcal Z_{\vec k}(t)\) are normalization constants.
The variance of the field -- the observable we focus on -- is given by
\begin{equation}
\mathcal{S}[\varphi]=\int_Vd\vec x\ \varphi^2(\vec x,t)=\frac{1}{V}\sum_{\vec k}\varphi_{\vec k}(t)\varphi_{-\vec k}(t).
\label{eq:samplevariance}
\end{equation}

The probability distribution of this quantity, at any time, is given by Eq.~(\ref{eq:initialpdf}) where 
the parameters \(\bf t\) amount to the time $t$ elapsed after the quench.
It has been shown in Ref.~\cite{CORBERI2015} that, for spatial 
dimensions \(d>2\), the probability distribution of this observable
is singular, in the sense that there exists a critical value of the fluctuations \(S_c(t)\)  
above which fluctuations are condensed. In particular, 
for \(S>S_c(t)\) the contribution to $P(S,t)$ of the zero mode in Eq.~(\ref{eq:samplevariance})
is macroscopic~\cite{Zannetti14}.
This implies that the rate function (see Eq.~(\ref{largeDevPr})) is strictly 
convex for \(s<s_c(t)\) while rectifies for \(s>s_c(t)\)~\cite{Zannetti14}, i.e.,
\begin{equation}
I(s,t)=\left \{ \begin{array}{ll}
z^*(s,t)s+\lambda(z^*(s,t),t) &  \mbox{for}\,\,\,\,\, s\le s_c(t) , \\
 (s-s_c)V/2\langle \varphi^2_{0}(t) \rangle+I(s_c,t) &  \mbox{for}\,\,\,\,\, s>s_c(t) .
\end{array}
\right .
\label{RateCase}
\end{equation}
Note that the slope of the straight part of $I$ depends only on the zero mode $\varphi _0$, 
because this is the condensing degree of freedom.

\section{Dynamics of fluctuations} \label{fluctuations}

Before discussing the quench dynamics, it is useful to recall a scaling property of the equilibrium state
and of the associated probability distribution \(P_{eq}(S)\)
that will be useful in the following. In particular, it turns out that 
\be
P_{eq}(S)=f\left ( \frac{S}{\langle S\rangle}\right ),
\label{scaleq}
\ee
where $\langle S \rangle =\int _0 ^\infty dS \, S\, P(S)=\beta ^{-1}\sum _{\vec k} \omega _k ^{-1}$ is the average 
value of $S$.
Indeed by changing variable as $\psi_{\vec k}= (\langle S(t)\rangle/V)^{-1/2}\varphi_{\vec k}$ in Eq.~(\ref{eq:initialpdf})
leads to
\be
\begin{array}{ll}
P(S,t)=\mathcal Z^{-1}(t)\int \mathcal D\psi \, \exp \left \{-\frac{1}{2V}
  \sum _{\vec k} \frac{\psi_{\vec k}\psi_{-\vec k}}
       {\langle \psi_{\vec k}(t)\psi_{-\vec k}(t)\rangle}\right \}\times
       \delta \left (\frac {1}{V}\sum _{\vec k}\psi_{\vec k}\psi_{-\vec k}
       -\frac{S}{\langle S(t)\rangle}\right ),
\end{array}
\label{scalneq}
\ee
where \(\mathcal Z\) is a normalization constant.       
At equilibrium the time dependencies disappear, and,
due to the presence of \(\beta\)
in $\langle S\rangle$, $\langle \psi_{\vec k}\psi_{-\vec k}\rangle $ is temperature independent.
Accordingly, if one measures $S$ in units of $\langle S \rangle$, plotting $P_{eq}(S)$ for
different values of the temperature, in equilibrium one observes collapse onto a single mastercurve.

\begin{figure}[h]
  \centering
\rotatebox{-90}{\resizebox{.5\textwidth}{!}{\includegraphics{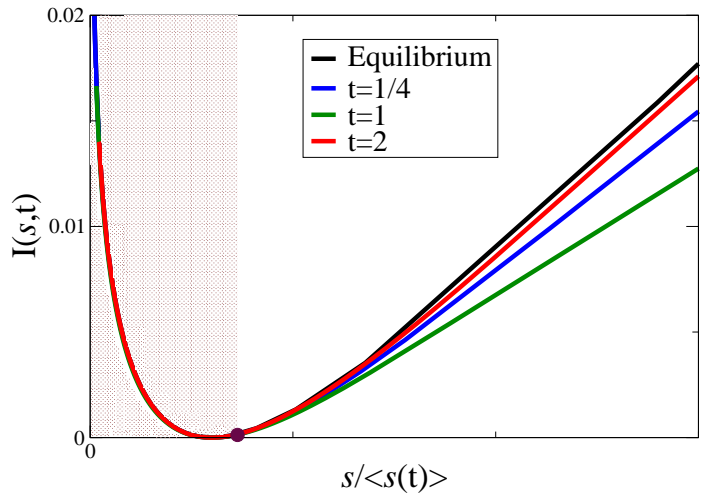}}}
  \caption{Rate function $I(s,t)$ for $r=1$ and $d=3$ plotted versus the rescaled variable $s/\langle s(t)\rangle$,
    for various times $t$ after a quench from $\beta _i=1/5$
    to $\beta _f=1$. The plots for \(t=0\) and \(t=\infty\) are represented by the same black line.
    The value $s_{inf}/\langle s(t)\rangle$ is marked by a
    thick dot while the brown background highlights the region where condensation never occurs.}
\label{fig}
\end{figure}

In Fig.~\ref{fig} the rate function \(I\), given by Eq.~(\ref{RateCase}), is plotted versus \(s/\langle s(t) \rangle\)
for the initial equilibrium state, for intermediates times of the quench dynamics and for the final state
reached by the system at the end of the equilibration process, in spatial dimension \(d=3\).
Due to the scaling law in Eq.~(\ref{scaleq}) the profile of the rate function for the system equilibrated
at the initial inverse temperature \(\beta_i\) and at the final one \(\beta_f\) is the same.
The solution of the model shows that the singular point $s_c(t)$, starting from its initial value \(s_c(0)\), 
decreases after the quench, reaches a minimum $s_{inf}=\inf _t [s_c(t)]$ at some time (the dot in Fig.~\ref{fig}), 
and then increases up to the final value
$s_c(\infty)<s_c(0)$. 

In order to discuss the non-equilibrium behavior of the rate function,  it is useful to separate the region
that is never interested by condensation $0\le s\le s_{inf}$, from the one with   $s> s_{inf}$,
where condensation plays a role. In fact, it is clearly seen in Fig.~\ref{fig}, that the evolution of the rate function is markedly different within these two regions. 
In the range of $s$ where condensation does not occur, the rate function overlaps with the equilibrium expression
at all times. This shows that fluctuations in this case behave as adiabatically in equilibrium at some 
time-dependent temperature corresponding to the actual value of $\langle s (t)\rangle $. 
The evolution of the fluctuations associated to condensed configurations for $s>s_{inf}$ is, instead, more complex.
Here the rate function deviates appreciably from the equilibrium form, from the very beginning of the 
non-equilibrium evolution. The branch of $I(s,t)$ with $s>s_c(t)$ is linear at all times, 
due to the condensation, but with a slope which depends on time. Specifically, its behavior follows
that of $s_c(t)$, i.e., it initially decreases and then, when $s_c(t)$ reaches $s_{inf}$, increases
until it attains the slope of the equilibrium state.
Note that, since smaller values of $I$ correspond to a larger probability of the corresponding fluctuations, 
this signals that out of equilibrium, the chances for the system to reach a configuration
affected by condensation increase, as already pointed out in Ref.~\cite{Zannetti14}.

The qualitatively different behavior occurring to the left and to the right of $s_{inf}$
can be heuristically explained as follows. The system stays adiabatically close to the equilibrium state
characterized by the actual value of $\langle s(t)\rangle$ if its own relaxation time is comparable
to the time associated with the evolution of $\langle s(t)\rangle$.
In the non-condensed phase the sample variance takes comparable contributions
from all the $k$-modes. Then we argue that the relaxation time of the system is comparable to the 
average relaxation time of all these modes. Solving the model it can be shown that this latter time is also 
comparable to the time associated with the displacement of \(\langle s(t) \rangle\). This explains why
the equilibrium collapse of Eq.~(\ref{scaleq}) is observed also out of equilibrium for $s\le s_{inf}$. 
For \(s>s_{inf}\) the contribution of the mode with $k=0$ becomes macroscopic 
and dominates the relaxation process. Since this is also the slowest mode (see Eq.~(\ref{eq:solution_k-modes})), the
overall evolution is much slower than that of $\langle s(t)\rangle$ and therefore the condition of adiabaticity 
is not fulfilled.

\section{Conclusions} \label{conclusions}

In this paper we have considered the Gaussian model of statistical mechanics, 
focusing on the fluctuations of the order parameter variance $\mathcal S$. 
Its probability distribution $P(S,t)$ displays a singular behavior due to the
phenomenon of condensation of fluctuations.

We have studied the evolution of $P(S,t)$, after a sudden temperature quench.
Compared to Ref.~\cite{2019arXiv190508536C}, where a kinetics with conservation of the order parameter 
was considered, here we have studied the dynamics without conservation, i.e., the so-called Model A~\cite{Hohenberg77}.
By solving exactly the evolution equations of the model we have shown that the dynamics of the fluctuations
is very different depending on the value of $S$ considered. Fluctuations associated with a value of $S$ which, during
the evolution, is not crossed by the evolution of the singular point $S_c(t)$ evolve in a fast and rather trivial way:
$P(S,t)$ always stays close to an equilibrium form, the only appreciable evolution being the shift of 
the typical value $\langle S(t)\rangle$, due to the system cooling. Because of this, fluctuations within this range
(sketched in Fig.~\ref{fig}) can be viewed as adiabatically equilibrated at a decreasing effective temperature.
On the other hand, fluctuations within a range at values of $S$ crossed by $S_c(t)$ during the evolution have a much slower and
complex dynamics.

This is due to the prominent role played by the contribution $s_0(s,t)$ provided by the $k=0$ mode which,
starting microscopic, has to grow macroscopically large.
The emergence of two different behaviors was already
observed in the case considered in Ref.~\cite{2019arXiv190508536C} and 
in another solvable mode in Ref.~\cite{Corberi17}, promoting this feature to a rather generic property.
Accordingly, although we have restricted our attention here to a temperature quench, 
we expect a similar scenario to be observed if other kind of quenches, e.g., a quench in the parameter $r$, 
were studied, as well as if other observables beyond the order parameter variance were considered.

As a final comment, we point out that the models where this kind of problems have been studied insofar are
such that fluctuating modes can be made independent by a suitable diagonalization.
However, singular probability distributions have been observed also in more complex and fully interacting
systems, for instance
in the height statistics of a growing interface~\cite{PhysRevE.96.020102,PhysRevE.97.042130} or
in intrinsically non-equilibrium states of active matter~\cite{Cagnetta17,Nemoto19},
where an analysis as the one carried out in this paper has not yet been conducted and remains a topic for future research.

\section*{Acknowledgments}

F.C. acknowledges funding from PRIN 2015K7KK8L.

\section*{References}

\bibliographystyle{iopart-num}

\bibliography{fluctuations}

\providecommand{\newblock}{}
\begin{thebibliography}{10}
\expandafter\ifx\csname url\endcsname\relax
  \def\url#1{{\tt #1}}\fi
\expandafter\ifx\csname urlprefix\endcsname\relax\def\urlprefix{URL }\fi
\providecommand{\eprint}[2][]{\url{#2}}

\bibitem{Hinrichsen00}
Hinrichsen H 2000 {\em Adv. Phys.\/} {\bf 49} 815

\bibitem{Langer92}
Langer J 1992 {\em Solids far from Equilibrium\/} (Cambridge: Cambridge
  University Press) pp 297--363

\bibitem{Touchette2009}
Touchette H 2009 {\em Phys. Rep.\/} {\bf 478} 1

\bibitem{Corberi19}
Corberi F and Sarracino A 2019 {\em Entropy\/} {\bf 21} 312

\bibitem{Baek_2015}
Baek Y and Kafri Y 2015 {\em J. Stat. Mech.\/} {\bf 2015} P08026

\bibitem{Filiasi_2014}
Filiasi M, Livan G, Marsili M, Peressi M, Vesselli E and Zarinelli E 2014 {\em
  J. Stat. Mech.\/} {\bf 2014} P09030

\bibitem{Harris_2009}
Harris R~J and Touchette H 2009 {\em J. Phys. A: Math. Theor.\/} {\bf 42}
  342001

\bibitem{Gradenigo_2013}
Gradenigo G, Sarracino A, Puglisi A and Touchette H 2013 {\em J. Phys. A: Math.
  Theor.\/} {\bf 46} 335002

\bibitem{Gambassi2012}
Gambassi A and Silva A 2012 {\em Phys. Rev. Lett.\/} {\bf 109} 250602

\bibitem{2019arXiv190406259P}
{Perfetto} G, {Piroli} L and {Gambassi} A 2019  arXiv:1904.06259

\bibitem{Goold2018}
Goold J, Plastina F, Gambassi A and Silva A 2018 {\em Thermodynamics in the
  Quantum Regime: Fundamental Aspects and New Directions\/} (Cham: Springer
  International Publishing) pp 317--336

\bibitem{Touchette2007}
Touchette H and Cohen E~G~D 2007 {\em Phys. Rev. E\/} {\bf 76} 020101

\bibitem{Touchette_2009}
Touchette H and Cohen E~G~D 2009 {\em Phys. Rev. E\/} {\bf 80} 011114

\bibitem{Bouchet_2012}
Bouchet F and Touchette H 2012 {\em J. Stat. Mech.\/} {\bf 2012} P05028

\bibitem{Harris_2005}
Harris R~J, R{\'{a}}kos A and Schütz G~M 2005 {\em J. Stat. Mech.\/} {\bf
  2005} P08003

\bibitem{Szavits2014}
Szavits-Nossan J, Evans M~R and Majumdar S~N 2014 {\em Phys. Rev. Lett.\/} {\bf
  112} 020602

\bibitem{Chleboun2010}
Chleboun P and Grosskinsky S 2010 {\em J. Stat. Phys.\/} {\bf 140} 846

\bibitem{Janas2016}
Janas M, Kamenev A and Meerson B 2016 {\em Phys. Rev. E\/} {\bf 94} 032133

\bibitem{Sasorov_2017}
Sasorov P, Meerson B and Prolhac S 2017 {\em J. Stat. Mech.\/} {\bf 2017}
  063203

\bibitem{Majumdar_2014}
Majumdar S~N and Schehr G 2014 {\em J. Stat. Mech.\/} {\bf 2014} P01012

\bibitem{langer2000theory}
Langer J~S 1967 {\em Ann. Phys. (N. Y.)\/} {\bf 41} 108

\bibitem{Zannetti14}
Zannetti M, Corberi F and Gonnella G 2014 {\em Phys. Rev. E\/} {\bf 90} 012143

\bibitem{CORBERI2015}
Corberi F, Gonnella G and Piscitelli A 2015 {\em J. Non-Cryst. Solids\/} {\bf
  407} 51

\bibitem{Corberi17}
Corberi F 2017 {\em Phys. Rev. E\/} {\bf 95} 032136

\bibitem{Cagnetta17}
Cagnetta F, Corberi F, Gonnella G and Suma A 2017 {\em Phys. Rev. Lett.\/} {\bf
  119} 158002

\bibitem{Corberi_2015}
Corberi F 2015 {\em J. Phys. A: Math. Theor.\/} {\bf 48} 465003

\bibitem{Zannetti_2014}
Zannetti M, Corberi F, Gonnella G and Piscitelli A 2014 {\em Commun. Theor.
  Phys.\/} {\bf 62} 555

\bibitem{Corberi_2013}
Corberi F, Gonnella G, Piscitelli A and Zannetti M 2013 {\em J. Phys. A: Math.
  Theor.\/} {\bf 46} 042001

\bibitem{Corberi_2012}
Corberi F and Cugliandolo L~F 2012 {\em J. Stat. Mech.\/} {\bf 2012} P11019

\bibitem{2019arXiv190508536C}
{Corberi} F, {Mazzarisi} O and {Gambassi} A 2019  {\em J. Stat. Mech.\/} {\bf 2019} P104001

\bibitem{Goldenfeld92}
Goldenfeld N 1992 {\em Lectures on Phase Transitions and the Renormalization
  Group\/} (Reading: Addison-Wesley)

\bibitem{chaikin_lubensky_1995}
Chaikin P~M and Lubensky T~C 1995 {\em Principles of Condensed Matter
  Physics\/} (Cambridge: Cambridge University Press)

\bibitem{Hohenberg77}
Hohenberg P~C and Halperin B~I 1977 {\em Rev. Mod. Phys.\/} {\bf 49} 435

\bibitem{PhysRevE.96.020102}
{Krajenbrink} A and {Le Doussal} P 2017 {\em Phys. Rev. E\/} {\bf 96} 020102

\bibitem{PhysRevE.97.042130}
{Smith} N~R, {Kamenev} A and {Meerson} B 2018 {\em Phys. Rev. E\/} {\bf 97}
  042130

\bibitem{Nemoto19}
Nemoto T, Fodor E, Cates M~E, Jack R~L and Tailleur J 2019 {\em Phys. Rev. E\/}
  {\bf 99} 022605

\end{thebibliography}

\end{document}